\documentclass[%
 reprint,
superscriptaddress,
nofootinbib,
nobibnotes,
 amsmath,amssymb,
 aps,
prb,
]{revtex4-1}

\usepackage{graphicx}
\usepackage{dcolumn}
\usepackage{braket}
\usepackage{gensymb}
\usepackage{bm}
\usepackage{amsmath}
\usepackage{float}
\usepackage{color}
\usepackage{units}
\usepackage{lipsum}
\usepackage{csquotes}
\usepackage[makeroom]{cancel}
\usepackage{multirow}
\usepackage{booktabs}

\usepackage[normalem]{ulem}

\usepackage{siunitx}
\usepackage{xcolor}

\usepackage{sidecap}

\makeatletter
\newcommand*{\rom}[1]{\expandafter\@slowromancap\romannumeral #1@}
\makeatother

\usepackage[utf8]{inputenc}

\begin{document}

\preprint{APS/123-QED}

\title{Genuine Counterfactual Communication with a Nanophotonic Processor}

\author{I. Alonso Calafell}
\thanks{These authors contributed equally to this work.\\
irati.alonso.calafell@univie.ac.at, teodor.stroemberg@univie.ac.at}
\author{T. Str\"{o}mberg}
\thanks{These authors contributed equally to this work.\\
irati.alonso.calafell@univie.ac.at, teodor.stroemberg@univie.ac.at}
\affiliation{Vienna Center for Quantum Science and Technology (VCQ), Faculty of Physics, University of Vienna, Boltzmanngasse 5, Vienna A-1090, Austria}
\author{D. R. M. Arvidsson-Shukur}
\affiliation{Cavendish Laboratory, Department of Physics, University of Cambridge, CB3 0HE, Cambridge, United Kingdom}
\author{L. A. Rozema}
\affiliation{Vienna Center for Quantum Science and Technology (VCQ), Faculty of Physics, University of Vienna, Boltzmanngasse 5, Vienna A-1090, Austria}
\author{V. Saggio}
\author{C. Greganti}
\affiliation{Vienna Center for Quantum Science and Technology (VCQ), Faculty of Physics, University of Vienna, Boltzmanngasse 5, Vienna A-1090, Austria}
\author{N. C. Harris}
\author{M. Prabhu}
\author{J. Carolan}
\affiliation{Quantum Photonics Group, RLE, Massachusetts Institute of Technology, Cambridge, Massachusetts 02139, USA}
\author{M. Hochberg}
\author{T. Baehr-Jones}
\affiliation{Elenion Technologies, New York, NY 10016, USA}
\author{D. Englund}
\affiliation{Quantum Photonics Group, RLE, Massachusetts Institute of Technology, Cambridge, Massachusetts 02139, USA}
\author{C. H. W. Barnes}
\affiliation{Cavendish Laboratory, Department of Physics, University of Cambridge, CB3 0HE, Cambridge, United Kingdom}
\author{P. Walther}
\affiliation{Vienna Center for Quantum Science and Technology (VCQ), Faculty of Physics, University of Vienna, Boltzmanngasse 5, Vienna A-1090, Austria}

\date{\today}


\begin{abstract}
In standard communication information is carried by particles or waves\cite{Shannon48,Bennett92,Schumacher95}. 
Counterintuitively, in counterfactual communication particles and information can travel in opposite directions. The quantum Zeno effect allows Bob to transmit a message to Alice by encoding information in particles he never interacts with \cite{Degasperis74, Misra77}. The first suggested protocol\cite{Salih13} not only required thousands of ideal optical components, but also resulted in a so-called ``weak trace''\cite{Danan13} of the particles having travelled from Bob to Alice, calling the scalability and counterfactuality of previous proposals\cite{Vaidman13,Vaidman13-2, Vaidman14, Griffiths16, ArvShukur17} and experiments \cite{Liu17, Cao17} into question. Here we overcome these challenges, implementing a new protocol \cite{ArvShukur16} in a programmable nanophotonic processor, based on reconfigurable silicon-on-insulator waveguides that operate at telecom wavelengths \cite{Harris2017}.
This, together with our telecom single-photon source and highly-efficient superconducting nanowire single-photon detectors, provides a versatile and stable platform for a high-fidelity implementation of genuinely trace-free counterfactual communication, allowing us to actively tune the number of steps in the Zeno measurement, and achieve a bit error probability below $\bm{1 \, \%}$, with neither post-selection nor a weak trace. Our demonstration shows how our programmable nanophotonic processor could be applied to more complex counterfactual tasks and quantum information protocols \cite{Noh09, Yin10, Liu14}.
\end{abstract}

\maketitle 

Interaction-free measurements allow one to measure whether or not an object is present without ever interacting with it\cite{dicke81}.
This is made clear in Elitzur and Vaidman's well-known bomb-testing gedanken experiment\cite{Elitzur93}.
In this experiment, a single photon used in a Mach-Zehnder interferometer (MZI) sometimes reveals whether or not an absorbing object (e.g. a bomb) had been placed in one of the interferometer arms, without any interaction between the photon and the bomb. 
It was later shown that the quantum Zeno effect, wherein repeated observations prevent the system from evolving \cite{Degasperis74, Misra77}, can be used to bring the success probability of this protocol arbitrarily close to unity \cite{Degasperis74, Misra77,Kwiat95,Kwiat99}.
Such protocols are often referred to as ``counterfactual'', and have now been applied to quantum computing\cite{Hosten06}, quantum key distribution \cite{Noh09, Yin10, Liu14} and communication \cite{Salih13, ArvShukur16}. Here we experimentally implement a counterfactual communication (CFC) protocol where information can propagate without being carried by physical particles.

\begin{figure}[h!]
\centering
\includegraphics[width=\columnwidth]{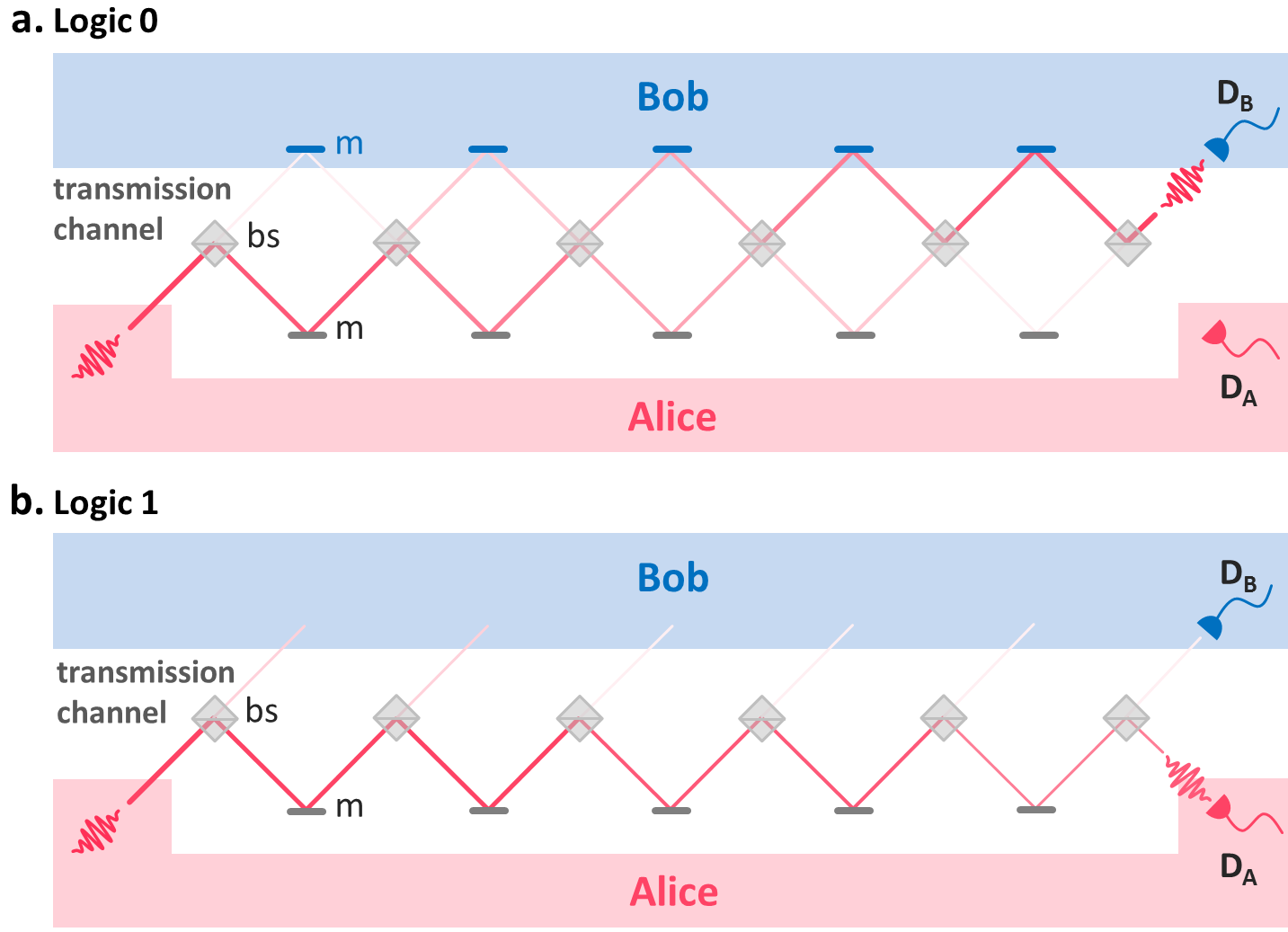}
\caption{\textbf{Architecture of the chained MZI protocol.} Alice inputs a photon into the transmission channel, consisting of a row of beamsplitters (BSs) and the lower row of mirrors (marked with an `m'). \textbf{a.} If Bob intends to send a logic 0, he places mirrors in his laboratory to form MZIs that span his lab and the transmission channel, creating constructive interference in Bob's port ($D_B$). \textbf{b.} If he intends to send a logic 1, he removes the mirrors, causing the photons to arrive back in Alice's laboratory ($D_A$) with high probability.}
\label{fig:HilbSpa}
\vspace{-4mm}
\end{figure}

The first suggested protocol for CFC was developed by Salih \textit{et al.}, and it is based on a chain of nested MZIs \cite{Salih13}. There are four main concerns with this scheme: (1) Achieving a high success probability (say $>\SI{95}{\percent}$) requires thousands of optical elements. (2) An analysis of the Fisher information flow shows that to retain counterfactuality in Salih's protocol, absolutely perfect quantum channels are needed\cite{ArvShukur17}. (3) Alice's particles leave a weak trace in Bob's laboratory, raising doubts about the scheme's counterfactuality\cite{Vaidman13, Vaidman13-2, Vaidman14}. (4) This scheme requires post-selection to remove the CFC violations.   


\begin{figure*}[t!]
\centering
\includegraphics[width=1\textwidth]{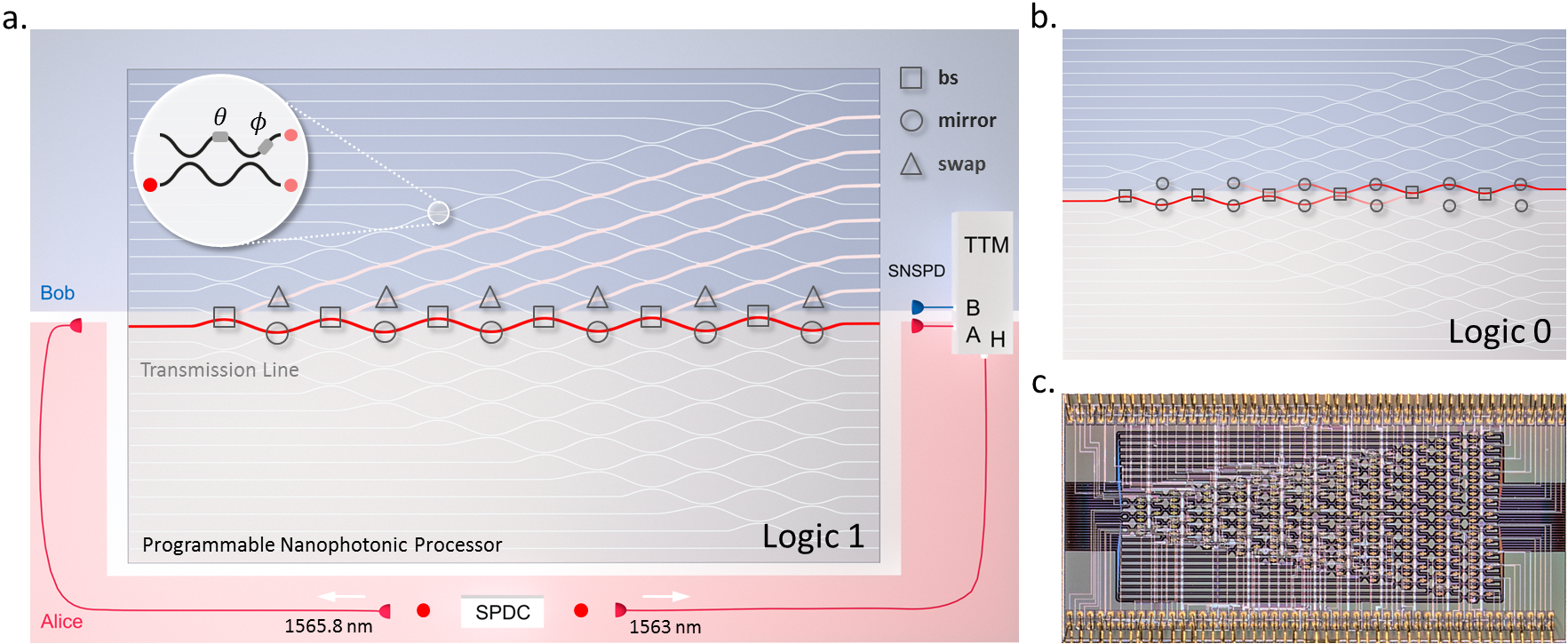}
\caption{\textbf{Experimental setup.} \textbf{a.} Our experiment is implemented in a programmable nanophotonic processor (PNP), which is composed of 26 interconnected waveguides. The waveguides are coupled by 88 Mach-Zehnder interferometers (MZIs), as indicated by the top-left inset.
Each MZI is equipped with a pair of thermo-optic phase shifters, which allows us to treat them as beamsplitters with fully tunable reflectivities (set via $\theta\in[0,2\pi]$) and phases (($\phi\in[0,2\pi]$). In our work, we set $\theta$ to $\pi$, $0$ or $\pi/2N$, to implement mirrors (circles), SWAPs (triangles) and beamsplitters (squares), respectively. 
In Alice's laboratory (the pink shaded region) a spontaneous parametric down-conversion source creates a frequency non-degenerate photon pair at $\lambda_H=\SI{1563}{\nano\meter}$, $\lambda_T=\SI{1565.8}{\nano\meter}$. 
Detection of the $\lambda_H$ photon in detector H heralds the $\lambda_T$ photon that is injected into the transmission channel. 
This channel is comprised of the lower half of the PNP, in which MZIs are set to act as mirrors, as well as the MZIs that couple the upper and lower half of the waveguide. The latter of these MZIs are configured to act as beamsplitters, whose reflectivity varies with $N$ (the number of beamsplitters used in the protocol) as $R(N)=\cos^2(\pi/2N)$. 
Bob's laboratory consists of the upper set of MZIs (blue shaded area), which he can set as mode swaps to send a logic $1$ or \textbf{b.} as mirrors to send a logic $0$.
The photons are detected by superconducting nanowire single-photon detectors with detection efficiencies of approximately $\SI{90}{\percent}$. Coincident detection events are recorded with a custom-made Time Tagging Module (TTM). \textbf{c.} Micrograph of the PNP with dimensions $\SI{4.9x2.4}{\milli\meter}$.}
\label{fig:chip}
\end{figure*}

To get around these issues, we implement a novel CFC protocol\cite{ArvShukur16} that does not need post-selection and requires orders of magnitude fewer optical elements than nested MZI protocols. In this scheme photons travel from Alice to Bob but information from Bob to Alice. Note that the very recent proposals\cite{Aharanov18,rarity} discussing means of making the Salih scheme trace-free still require the post-selected removal of non-counterfactual events as well as thousands of ideal optical components. 

We perform our experiment using telecom single-photons in a state-of-the-art programmable nanophotonic processor (PNP) \cite{Harris2017}, which is orders of magnitude more precise and stable than previous bulk-optic approaches \cite{Kwiat95, Kwiat99}. Our PNP also provides unprecedented tunability, which we use to investigate the scaling of the protocol by changing the number of chained interferometers. By combining the novel CFC protocol with our advanced photonic technology, we are able to implement counterfactual communication with a bit success probability above \SI{99}{\percent}, without post-selection. 

Our protocol uses a series of $N$ beamsplitters with reflectivity $R=\cos^2(\pi/2N)$, which, together with mirrors, form a circuit of $N-1$ chained MZIs. As shown in Figure \ref{fig:HilbSpa}, the communication protocol begins with Alice injecting a single photon into her input port. If Bob wants to send a logic $0$ he leaves his mirrors in place, causing the photon to self interfere such that it exits in $D_B$ with unit probability (Figure \ref{fig:HilbSpa}a). To send a logic $1$ Bob \textit{locally} modifies the circuit to have the upper paths open (Figure \ref{fig:HilbSpa}b). In this case the photon will successfully reflect off of all the beamsplitters and exit in $D_A$ with probability $R^N$. Removing the mirrors effectively collapses the wavefunction after every beamsplitter, suppressing interference and implementing the Zeno effect. The probability that the photon remains in the lower arm after $N$ beamsplitters can be made arbitrarily high by increasing $N$ (and changing the reflectivities accordingly). 
%
%
\begin{figure*}[t!]
\centering
\includegraphics[width=1\textwidth]{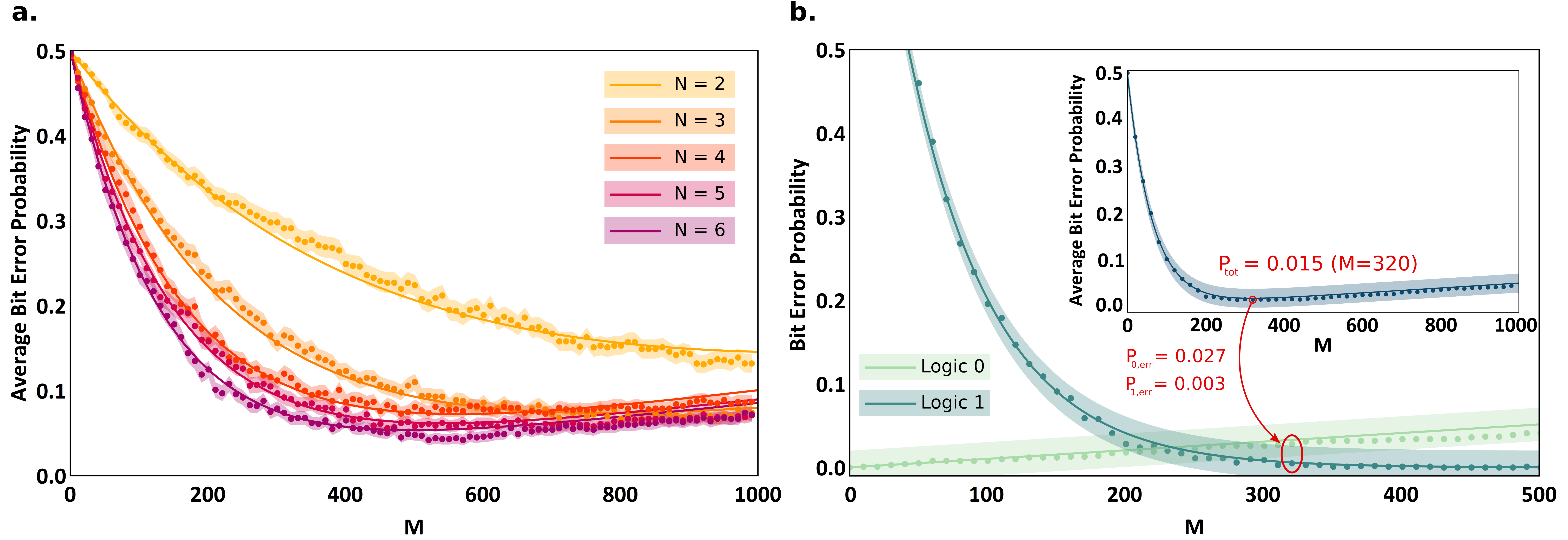}
\caption{\textbf{Success probabilities of the CFC communication.} The curves are theoretical models of our experiment with no free parameters, and the points are experimental data. \textbf{a.} Measured average bit error (as defined in the main text) of the protocol for different number of beamsplitters ($N$) as a function of the number of photons (\textit{M}) used to encode each bit. For small $M$ the $\cos^{2N}(\pi/2N)$ dependence of the logic $1$ error dominates the average error, making the latter decrease with $M$ as expected. As $M$ is increased more, the linearly growing error in the logic $0$, caused by imperfect destructive interference in Alice's port ($D_A$), starts to dominate.
\textbf{b.} In the $N=6$ case, the optimization of the interferometer fidelity and heralding efficiency leads to an average bit error rate of \SI{1.5}{\percent} for $M=320$, where the average CFC violation probability is \SI{2.4}{\percent}.} 
\label{fig:Ns}
\end{figure*}
%
%
%
%

Since any implementation is restricted to a finite number of beamsplitters, there will be a probability for a photon to exit the wrong port when Bob tries to send a logic $1$. This error probability is a function that decreases with $N$ as $P_{1,err}=1-R(N)^N$. In the non-ideal case, optical losses in the system will increase this probability further. The errors associated with Bob's attempt to transmit a logic $0$ are of a different nature. In theory, he can always perfectly transmit a logic $0$, independent of $N$; that is, $P_{0,err}=0$. In practice, however, imperfections in the interferometers will lead to cases in which the photon re-enters Alice's laboratory and she incorrectly records a logic $1$. This leads to a counterfactual violation, as the wavefunction ``leaks'' from Bob's to Alice's laboratory\cite{ArvShukur17}. Although they do not contribute to a counterfactual violation, dark counts in Alice's detector will also increase this error rate.

We can overcome the bit errors by encoding each logical bit into $M$ single photons, at the cost of slightly increasing the CFC violation. If Alice sends $M$ photons into the transmission channel without detecting any at $D_A$, she will record a logic $0$. On the other hand, if she detects one or more photons in her laboratory, she will record a logic $1$. Assuming messages with an equal number of $0$s and $1$s, the average bit error probability is given by:
\begin{equation}
\overline{P}_{err}(M) = \frac{1}{2} \big[  (P_{1,err})^M + MP_{0,err}   \big]
\end{equation} 
where the second term is an approximation of $1-P_0^M$ valid for small values of $MP_{0,err}$. By increasing $M$ we can thus decrease the contributions of $P_{1,err}$ exponentially while only increasing those of $P_{0,err}$ linearly. The counterfactual violation probability for a random bit is given by 
\begin{equation}
\overline{P}_{CFC}(M) = \frac{1}{2\eta} MP_{0,err},
\end{equation} 
where $\eta$ is the detector efficiency. We can thus find an $M$ that minimises the average bit error, while also maintaining a low counterfactual violation probability. In our experiment this expression slightly overestimates the violation probability, as it includes the detector dark counts. 

As illustrated in Figure \ref{fig:chip}, we implement a series of cascaded MZIs using a PNP.
At the intersections of each of the modes shown in the figure there are smaller MZIs that act as beamsplitters with tunable reflectivities and phases. 
This allows us to vary our CFC protocol using two to six beamsplitters.
Additionally, the high interferometric visibility of the PNP, which we measure to be \SI{99.94}{\percent} on average, allows us to keep the rate of counterfactual violations low.
The single photons are generated in a spontaneous parametric down conversion process and detected using superconducting nanowire single-photon detectors with detection efficiencies $\eta\sim\SI{90}{\percent}$ (see Appendix).
\begin{figure*}[!t]
\centering
\includegraphics[width=1\textwidth]{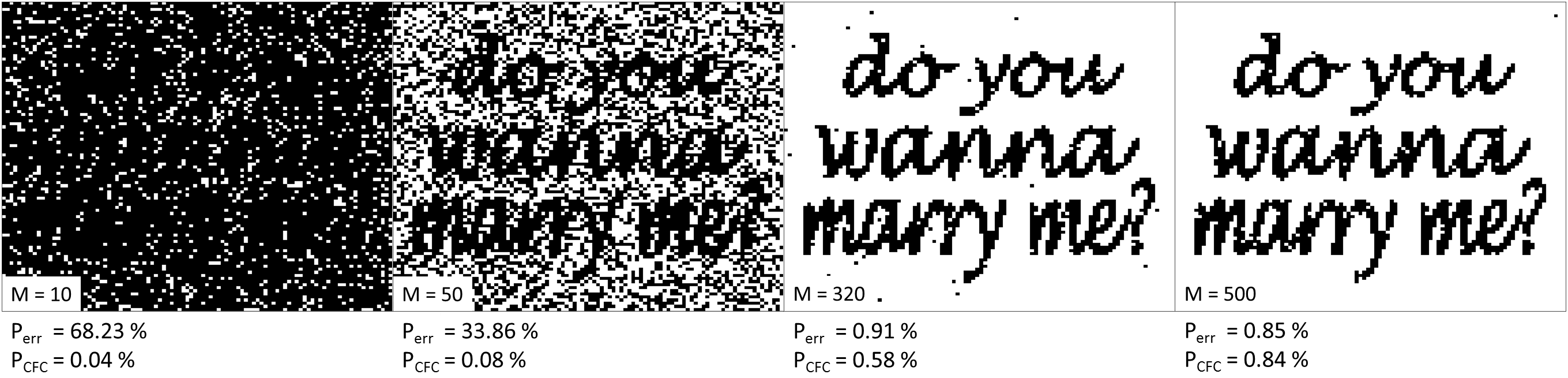}
\caption{Image sent from Bob to Alice by encoding bits in different numbers of single photons $M=\{10,50,320,500\}$. The white and black pixels are defined to correspond to logic 1 and logic 0, respectively. The success probability increases with increasing $M$, reaching \SI{99}{\percent} for $M=320$. The CFC violation probability ($P_{CFC}$) also increases with increasing $M$, but it remains as low as \SI{0.6}{\percent} for $M=320$. Note that this CFC violation comes only from the logic 0 errors, which we can directly measure; the total CFC violation is larger as discussed in the main text. Increasing $M$ beyond 320 increases the success probability at the expense of increasing the CFC violation. As it can be observed, these probabilities are directly related to the transmission fidelity ($F$) of the white pixels, which increases with $M$, and the transmission fidelity of the black pixels, which decreases with $M$. }
\label{fig:ImageM}
\end{figure*}  

To study the performance of this CFC protocol we measure the average bit error, as a function of the number of photons in which the bit is encoded, \textit{M}, for five different values of $N$ number of BSs. For the logic 0, we configure the MZIs in Bob's laboratory as mirrors (see Figure \ref{fig:chip}), while for the logic $1$ we let the MZIs in Bob's laboratory act as SWAP gates, routing the light out of the interferometer chain. Since Alice cannot access detector $D_B$, she assumes that a photon is injected in the transmission channel every time she detects a heralding photon in $D_H$. We thus run the measurement until we have \textit{M} recorded single-photon events in $D_H$ (typical rates were \SI{1.1}{\mega\hertz}) and look for the coincidences that these events have with $D_A$ within a set coincidence window $\Delta\tau=\SI{2.5}{\nano\second}$ that is shorter than the pulse separation. Our heralding efficiency was $\sim\SI{3}{\percent}$ through the PNP.

Figure \ref{fig:Ns}a shows the experimental average error probability of our CFC protocol as a function of $M$ for different $N$. We also include a theoretical calculation of the expected error probabilities, which considers the heralding efficiency of the single photons and the success probability of the interferometer that is in good agreement with the experimental data. Note that these are not fits to the data, but rather models with no free parameters. As theoretically predicted, the error rate of the logic 1 decreases exponentially with increasing $M$ and the error rate of the logic 0 increases linearly with $M$. We observe that higher $N$ requires smaller $M$, and also results in lower bit error probabilities.

The success probability of this CFC scheme is highly sensitive to the fidelity of the interferometers and the overall heralding efficiency, which depends on the single-photon source and the coupling efficiency throughout the system. Hence, we optimized the setup for the $N=6$ case. Figure \ref{fig:Ns}b shows the corresponding error probability of the logic $1$ and the logic $0$. The inset in Figure \ref{fig:Ns}b shows the average error probability, where we find a minimum of \SI{1.5}{\percent} for $M=320$, while the average counterfactual violation is kept at \SI{2.4}{\percent}. Owing to on-chip backscattering in Bob’s laboratory (i.e. imperfect SWAP operations) a small ``amount'' of wavefunction amplitude leaks back into the transmission line in the $1$ bit process. Although these do not all lead to detection events in Alice's laboratory, the sum of their squares provides an upper bound on the probability of a counterfactual violation. We estimate that the probability for a photon to reflect off of a SWAP operation is at most \SI{1}{percent}. Hence, in our experiment (Fig. \ref{fig:ImageM}) with $M=320$ and $N=6$, the contribution from the logic $1$ to a CFC violation is less than \SI{1.1}{\percent}. Note that this violation probability decreases with $N$, even if the errors remain the same.

To demonstrate the performance of the communication protocol we proceed to analyse the quality of a message in the form of a black and white image, sent from Bob to Alice, for $N=6$ and $M=\{10,50,320,500\}$. We arbitrarily define the white and black pixels of the image as logic $1$ and logic $0$, respectively. 

Figure \ref{fig:ImageM} shows the message transmitted from Bob to Alice for different numbers of encoding photons. We define the image fidelity as
\begin{equation}
F=\sum_{i=1}^{T}\frac{1+(-1)^{A_i+B_i}}{2T}
\end{equation}
where $B_i$ is the bit that Bob intended to send, $A_i$ is the bit that Alice recorded, and $T$ is the total number of bits in the image. In this case we define the CFC violation probability as the number of incorrectly transmitted logic $0$s (black pixels) over $T$.  
The encoding using $M=10$ is clearly not enough to overcome the losses of the system. As we increase $M$, the success probability and legibility of the message increases, reaching \SI{99}{\percent} for $M=320$, while the CFC violation probability from 0 bit errors remains as low as \SI{0.6}{\percent}. If the CFC violation of the 1 bit is accounted for, this value increases to \SI{2.3}{\percent}. For larger $M$s the success probability increases, but so does the CFC violation. Note that these values are lower than the value in Figure \ref{fig:Ns}b due to the unbalanced distribution of black and white pixels in the image.

Our high-fidelity implementation of a trace-free counterfactual communication protocol without post-selection was enabled by a programmable nano-photonic processor. The high (\SI{99.94}{\percent}) average visibility of the individual integrated interferometers allowed bit error probabilities as low as \SI{1.5}{\percent}, while, at the same time, keeping the probability for the transmission of a single bit to result in a counterfactual violation below \SI{2.4}{\percent}. By combining our state-of-the-art photonic technology with a novel theoretical proposal we contradicted a crucial premise of communication theory\cite{Shannon48}: that a message is carried by physical particles or waves. In fact, our work shows that ``interaction-free non-locality'', first described by Elitzur and Vaidman\cite{Elitzur93}, can be utilised to send information that is not necessarily bound to the trajectory of a wavefunction or to a physical particle. In addition to enabling further high-fidelity demonstrations of counterfactual protocols, our work highlights the important role that technological advancements can play in experimental investigations of fundamentals of quantum mechanics and information theory. We thus anticipate nanophotonic processors, such as ours, to be central to future photonic quantum information experiments all the way from the foundational level to commercialized products.

\begin{acknowledgements}
I.A.C. and T.S. acknowledge support from the University of Vienna via the Vienna Doctoral School. L.A.R. acknowledges support from the Templeton World Charity Foundation (fellowship no. TWCF0194). P.W. acknowledges support from the European Commission through ErBeSta (No. 800942), the Austrian Research Promotion Agency (FFG) through the QuantERA ERA-NET Cofund project HiPhoP,  from the Austrian Science Fund (FWF) through CoQuS (W1210-4) and NaMuG (P30067-N36), the U.S. Air Force Office of Scientific Research (FA2386-233 17-1-4011), and Red Bull GmbH.

D.R.M.A.S. acknowledges support from the EPSRC, Hitachi Cambridge, Lars Hierta's Memorial Foundation and the Sweden-America Foundation. N.H. was supported in part by the Air Force Research Laboratory RITA program 
(FA8750-14-2-0120); 
Research program FA9550-16-1-0391, supervised by Gernot Pomrenke; and D.E. acknowledges partial support from the Office of Naval Research CONQUEST program. 

The authors would like to express their gratitude towards J. Zeuner for helpful discussions and T. R\"{o}gelsperger for the artistic input. They furthermore thank LioniX International BV for the manufacturing of the interposer waveguides used in the experiment.
\end{acknowledgements}

\section*{Appendix}

\textit{Telecom Photon Source---} We use a pulsed Ti:Sapphire laser with a repetition rate of \SI{76}{\mega\hertz}, an average power of \SI{0.2}{\watt}, a central wavelength of \SI{782.2}{\nano\meter}, and a pulse duration of \SI{2.1}{\pico\second}. The repetition rate is doubled via a passive temporal multiplexing stage \cite{Greganti, Broome}. This beam pumps a periodically-poled KTP crystal phase matched for collinear type-II spontaneous parametric down conversion, generating frequency non-degenerate photon pairs at $\lambda_H=\SI{1563}{\nano\meter}$, $\lambda_T=\SI{1565.8}{\nano\meter}$. Registering the shorter wavelength photon at the detector $D_{H}$ heralds the presence of the longer wavelength one, which is sent to the waveguide.  

\textit{Programmable Nanophotonic Processor---} Our cascaded Mach-Zehnder interferometers (MZIs) are implemented in a silicon-on-insulator (SOI) programmable waveguide, developed by the Quantum Photonics Laboratory at the Massachusetts Institute of Technology \cite{Harris2017}. The device consists of 88 MZIs, each accompanied by a pair of thermo-optic phase shifters that facilitate full control over the internal and external phases of the MZIs. The phase shifters are controlled by a 240-channel, 16-bit precision voltage supply, allowing for a phase precision higher than \SI{250}{\micro\radian}. The coupling of the single photons in/out of the chip is performed using two $\textrm{Si}_3\textrm{N}_4\textrm{-SiO}_2$ waveguides manufactured by Lionix International, that adiabatically taper the \SI{10}{\micro\meter} x \SI{10}{\micro\meter} mode from the single mode fiber down to \SI{2}{\micro\meter} x \SI{2}{\micro\meter}, matching the mode field diameter of the programmable waveguide at the input facet. The total insertion loss per facet was measured as low as \SI{3}{\decibel}.

\textit{Superconducting Nanowire Single-Photon Detectors---} The photons are detected using superconducting nanowire single-photon detectors\cite{Natarajan,Marsili2013}. These detectors are produced by photonSpot and are optimized to reach detection efficiencies $\sim \SI{90}{\percent}$ at telecom wavelengths.

\bibliography{QuTexpVersionArXiv}

\end{document}